# Surface effects influencing the single-atom spontaneous emission in a linear atomic chain

Konstantin V. Krutitsky [*] and Jürgen Audretsch [†]

Fakultät für Physik, Universität Konstanz, Fach M 673, D-78457 Konstanz, Germany

**ABSTRACT**

As a contribution to quantum optics in the vicinity of surfaces we study the single atom spontaneous emission in a linear chain of two-level atoms. The electromagnetic field is thereby treated with the help of integro-differential equations which take into account the interaction with the other atoms in the chain. The life time of the excited atom, the frequency shift of the atomic transition and the angular distribution of emitted photons are worked out. They depend on the position of the emitting atom. As compared with the single atom in free space, considerable modifications occur for atoms a few interatomic distances away from the ends of the chain.

---

[*]Permanent address: Ulyanovsk Branch of Moscow Institute of Radio Engineering and Electronics of Russian Academy of Sciences, 48, Goncharov Str., Ulyanovsk 432700, Russia.

[†]e-mail: Juergen.Audretsch@uni-konstanz.de

# 1. INTRODUCTION

During the last decades there has been a growing interest in the behaviour of atoms and their spontaneous emission in particular near the surfaces of media. See, for example, the reviews [1]- [7]. The common method to solve boundary problems in quantum optics is based on the solution of Maxwell equations for the macroscopic electric field strength. Thereby the respective boundary conditions must be specified. The spatial geometry of the medium and the optical properties are given by the dielectric permittivity $\varepsilon$. The initially excited atom is characterized by the transition frequency and the transition dipole moment.

At present physical situations are in the center of the interest in which optical phenomena take place in the nearest vicinity of the media interfaces. At distances much less than the wave length of the spontaneous photons, various surface effects play an important role. For example, the influence of evanescent modes on an atom near a vacuum-dielectric interface has been studied in [8,9]. The atomic spectral characteristics and the spatial distribution of photons emitted spontaneously by the atom located in vacuum at small distances from the interface differ in principle from corresponding characteristics when the atom is located at large distances from the surface. In the theoretical analysis carried out in [8,9] the standard approach was used: the electromagnetic field was described by the macroscopic Maxwell equations and the medium was considered to be a continuous system of radiators.

The excistence of evanescent modes is not the only surface effect which may modify influence on the process of spontaneous emission. A different type of influence shows up if the *discrete* atomic structure of the medium is taken into account. In this case it can be demonstrated that even in a spatially uniform dielectric medium the optical properties of the surface layers differ significantly from the bulk properties. This is due to the violation of the translational symmetry at the surface. For a classical and a quantum optical treatment see [10]- [12] and [13,14], respectively. Here the characteristics of the spontaneous emission of an atom for the case of semi-infinite medium were obtained.

Regarding the description of the medium, in the papers [13,14] a very promising microscopic approach to the electromagnetic field was introduced which is based on integro-differential equations for the local field strength which is generated by a system of single radiators (atoms, molecules) situated at specific places with particular physical properties [15,16]. In the discussion below we will continue to apply the *local field approach* to the photon field but change to a medium with different geometry.

In recent years the investigation of the optical properties of linear molecular aggregates has obtained a lot of attention. In many cases such systems may be successfully modeled as linear chains of two-level atoms interacting by dipole-dipole interactions. At present the nonlinear optical properties of these systems [17,18] and the process of collective spontaneous emission (superradiation) [19] are sufficiently well understood. The results in papers [18,19] are again obtained on the basis of the classical electromagnetic theory. They indicate that in the radiation field the evolution of the atoms, located very close to the ends of the chain, is essentially different from the corresponding evolution of atoms located far away from the ends. This motivated the investigations of the present paper.

Below we study in the framework of the Weisskopf-Wigner approximation and based on the local field approach the process of spontaneous emission of a two-level atom embedded in a linear chain of $N-1$ ground state atoms. We will show that the spectral characteristics of the atom, as the life time of the excited state and the frequency shift of the atomic transition (Lamb shift), depend strongly on the atomic position if the atom is located very close to the ends of the chain while this is not the case at a larger distance. The angular distribution of the spontaneously emitted photons is also investigated. We thereby rely on the mathematical formalism developed in papers [13,20] but in contradistinction to the works [18,19] the electromagnetic field will be quantized.

The paper is organized as follows: In Section 2 the general formalism is presented which is based on the Heisenberg equations for the atomic operators and the field operators and an integro-differential equation for the local electromagnetic field strength. In Section 3 we calculate the electromagnetic mode functions for the linear atomic chain. Section 4 is devoted to the investigation of the life time and energy shift of an excited atom embedded in the linear chain. In Section 5 we investigate the angular distribution of the spontaneously emitted photons. Concluding remarks are made in Section 6.

## 2. THE FUNDAMENTAL EQUATIONS

We firstly describe a system of $N$ two-level atoms located at arbitrary positions $\mathbf{r}_l$ ($l = 1, ..., N$). We assume the atom with number $j$ ($j = 1, ..., N$) with transition frequency $\omega_0$ to be in the excited state while the other atoms with transition frequency $\omega \approx \omega_0$ are in their ground states. Our aim is to study the process of spontaneous emission emission of this atom. As compared with the free space case (one excited atom in vacuum) this process will be modified by the presence of the other atoms which are assumed to be polarizable and may therefore have an induced dipole moment. The initially excited atom which performs a dipole transition is during this process in a dipole-dipole interaction with all the other atoms. In our approach this will be represented by the adequately chosen vector potential operator. We then have for the initially excited atom at the position $\mathbf{r}_j$ the following effective Hamiltonian [21,22,13]:

$$H = \frac{1}{2}\hbar\omega_0\sigma_3 + \frac{\omega_0}{c}d_0\mathbf{u}_d\mathbf{A}(\mathbf{r}_j,t)\sigma_2 + \sum_{\mathbf{k}\lambda}\hbar\omega_k c^+_{\mathbf{k}\lambda}c_{\mathbf{k}\lambda} \quad , \tag{2.1}$$

where $\sigma_2$, $\sigma_3$ are the Pauli matrices. $\mathbf{u}_d = \mathbf{d}_0/d_0$ with $\mathbf{d}_0$ being the matrix element of the electric dipole moment operator is calculated with the aid of the unperturbed wave functions. $c^+_{\mathbf{k}\lambda}$, $c_{\mathbf{k}\lambda}$ are the photon creation and annihilation operators for the $\mathbf{k}\lambda$-mode. We assume that the interaction of the atom with the photon field isn't too strong, so that the Weisskopf-Wigner approximation holds.

The vector potential operator $\mathbf{A}(\mathbf{r}_j,t)$ needs a particular attention. At the position of the $j$-th atom it may be written in the Heisenberg representation in the form of the mode decomposition

$$\mathbf{A}(\mathbf{r}_j,t) = \sum_{\mathbf{k}\lambda} g'_{k\lambda}\mathbf{f}_{\mathbf{k}\lambda}(\mathbf{r}_j)c_{\mathbf{k}\lambda}(t) + h.c., \quad g'_{k\lambda} = \sqrt{\frac{2\pi\hbar c^2}{V_R\omega_k}}, \tag{2.2}$$

where $V_R$ is a quantization volume and $\mathbf{f}_{\mathbf{k}\lambda}$ is the mode function for the physical situation sketched above. It corresponds to the wave vector $\mathbf{k}$ and the polarization index $\lambda$ with $\omega_k = ck$.

In order to obtain the mode functions it is a common approach to solve Maxwell equations with appropriate boundary conditions. This procedure has the disadvantage that apart from evanescent waves [8] various surface effects are not taken into account. Because we are interested in the following in such effects which go back to the violation of the translational symmetry at the boundary of the medium, we choose a different approach which takes into account the physical properties of single atoms located at specific places. This method turns out to be particularly suited for the study of quantum optical effects which happen very close to the boundary. Instead of the macroscopic Maxwell equations we use thereby an integro-differential equation for the local classical field strength [15,16]. In the case of the discrete system of polarizable atoms it has the form of a system of coupled linear equations

$$\mathbf{f}_{\mathbf{k}\lambda}(\mathbf{r}_j) = \mathbf{e}_\lambda \exp(ik\mathbf{s}\mathbf{r}_j) + \sum_{l \neq j} rot\, rot_j \alpha_l \mathbf{f}_{\mathbf{k}\lambda}(\mathbf{r}_l) G(R_{lj}) \quad , \quad j = 1, ..., N, \qquad (2.3)$$

with $G(R) = \exp(ikR)/R$ and $R_{lj} = |\mathbf{r}_l - \mathbf{r}_j|$. Differentiation in (2.3) is carried out with respect to coordinates $\mathbf{r}_j$. $\alpha_l$ is the direction independent polarizability of the $l$-th atom. $\mathbf{e}_\lambda$ is the unit vector of polarization, $\mathbf{s} = \mathbf{k}/k$ is the unit vector in the direction $\mathbf{k}$ of the mode in question and $k$ is the wave number. Note that because of the second term in (2.3) which represents the influence of the other atoms ($l \neq j$) at the position $\mathbf{r}_j$ of the excited atom, the modulus $|\mathbf{f}_{\mathbf{k}\lambda}(\mathbf{r}_j)|$ will depend on $\mathbf{r}_j$. This will lead to the boundary effects which show up in modified characteristics of the process of spontaneous emission as compared with the free space case (only the first term in (2.3)). The solution of equations (2.3) will be discussed below.

Regarding the physical interpretation of (2.3) it has been shown in [20] that the quantum analogue of this integro-differential field equation may be obtained within the framework of third-order QED effects. Thereby the evolution of an $N$-atom system in the radiation field has been investigated in the dipole approximation. Each atom can emit or absorb one real photon independently of the others. In third order perturbation theory this process includes the interaction between the atoms through the exchange of virtual photons. The latter results in the second term of (2.3).

To work out the life time of the excited atom and the frequency shift of its atomic transition, we turn to the Heisenberg equations. Introducing the notations

$$A_j = \mathbf{A}\mathbf{u}_d = \sum_{\mathbf{k}\lambda} g_{\mathbf{k}\lambda}(\mathbf{r}_j) c_{\mathbf{k}\lambda} + h.c. \quad , \quad g_{\mathbf{k}\lambda}(\mathbf{r}_j) = g'_{\mathbf{k}\lambda} \mathbf{f}_{\mathbf{k}\lambda}(\mathbf{r}_j) \mathbf{u}_d \quad , \tag{2.4}$$

we obtain as in [23]

$$\dot{\sigma}_1 = -\omega_0 \sigma_2 + 2 \frac{\omega_0 d_0}{\hbar c} A_j \sigma_3 \quad ,$$

$$\dot{\sigma}_2 = \omega_0 \sigma_1 \quad , \tag{2.5}$$

$$\dot{\sigma}_3 = -2 \frac{\omega_0 d_0}{\hbar c} A_j \sigma_1 \quad ,$$

$$\dot{c}_{\mathbf{k}\lambda} = -i\omega_k c_{\mathbf{k}\lambda} - i \frac{\omega_0 d_0}{\hbar c} g^*_{\mathbf{k}\lambda}(\mathbf{r}_j) \sigma_2 \quad .$$

Equations (2.5) agree structurally with those for the spontaneous emission of a single atom in the free space. The solution for a not too short time $t$ after the beginning of the spontaneous emission is

$$c_{\mathbf{k}\lambda}(t) = c^{free}_{\mathbf{k}\lambda}(t) + i \frac{\omega_0 d_0}{\hbar c} g^*_{\mathbf{k}\lambda}(\mathbf{r}_j) \left[ \sigma^+(t) \zeta^*(\omega_k + \omega_0) - \sigma(t) \zeta^*(\omega_k - \omega_0) \right] \quad ,$$

$$\langle \sigma_3(t) \rangle = -1 + [1 + \langle \sigma_3(0) \rangle] \exp\left[-t/\tau(\mathbf{r}_j)\right] \quad , \tag{2.6a}$$

$$\langle \sigma^+(t) \rangle = \langle \sigma^+(0) \rangle \exp\left[i\left(\omega_0 + \delta(\mathbf{r}_j)\right) t\right] \exp\left[-t/2\tau(\mathbf{r}_j)\right] \quad ,$$

with

$$\frac{1}{\tau(\mathbf{r}_j)} = 2\pi \left(\frac{\omega_0 d_0}{\hbar c}\right)^2 \sum_{\mathbf{k}\lambda} |g_{\mathbf{k}\lambda}(\mathbf{r}_j)|^2 \left[\delta(\omega_k - \omega_0) - \delta(\omega_k + \omega_0)\right] \quad ,$$

$$\Delta(\mathbf{r}_j) = \left(\frac{\omega_0 d_0}{\hbar c}\right)^2 \sum_{\mathbf{k}\lambda} |g_{\mathbf{k}\lambda}(\mathbf{r}_j)|^2 \left(\frac{P}{\omega_k + \omega_0} - \frac{P}{\omega_k - \omega_0}\right) \quad . \tag{2.6b}$$

Here $\sigma^+$, $\sigma$ are the raising and lowering operators of the two-level atom, respectively; $\zeta(x)$ is a Heitler function [24]. Brackets $\langle \cdots \rangle$ denote vacuum averaging over the state $|\Psi\rangle = |vacuum\rangle|\Phi\rangle$, where $|\Phi\rangle$ is an arbitrary state of two-level atom and $|vacuum\rangle$ is the state without photons.

Regarding the interpretation of (2.6a) we see that the photon operator refers in its first term to the free space radiation field and in the additional term to the photon source part. The latter field is distorted by the medium (atoms with $l \neq j$) in accordance with the coordinate dependence of the mode functions $\mathbf{f}_{\mathbf{k}\lambda}(\mathbf{r}_j)$. Furthermore, it can directly be read off from (2.6a) that $\tau$ is the life time of the excited state and $\Delta$ the frequency shift of the atomic transition. Both now depend on the position $\mathbf{r}_j$ of the excited atom so that surface effects will become evident.

## 3. MODE FUNCTIONS OF A ONE-DIMENSIONAL CHAIN OF ATOMS

We specialize the situation described above and consider a one-dimensional straight chain of $N$ atoms with equal polarizability $\alpha$ located with interatomic distance $a_0$ on the negative part of the $z$-axis, so that $\mathbf{r}_l = (0, 0, z_l)$ with $z_l = -a_0(l-1)$ and $l = 1, \cdots, N$. Performing the differentiations in (2.3) we obtain the following system of equations for the mode functions:

$$\mathbf{f}_j = \mathbf{e}_\lambda \exp\left[-iks_z a_0(j-1)\right] + \tag{3.1}$$
$$+\alpha \sum_{l \neq j} \left\{ k^2 \frac{\mathbf{f}_l - f_l^z \mathbf{n}_z}{a_0|l-j|} - ik \frac{3f_l^z \mathbf{n}_z - \mathbf{f}_l}{a_0^2(l-j)^2} + \frac{3f_l^z \mathbf{n}_z - \mathbf{f}_l}{a_0^3|l-j|^3} \right\} \exp(ika_0|l-j|),$$

where $\mathbf{f}_l = \mathbf{f}_{\mathbf{k}\lambda}(\mathbf{r}_l) = (f_l^x, f_l^y, f_l^z)$. $\mathbf{n}_z$ is a unit vector in the direction of the $z$-axis. It is convenient to introduce spherical coordinates. The direction of the wave vector $\mathbf{k}$ of the mode becomes:

$$\mathbf{s} = (\sin\theta\cos\varphi, \sin\theta\sin\varphi, \cos\theta) \quad, \tag{3.2}$$

where $\theta$ is an incident angle which is counted from the positive direction of the axis $z$ ($0 \leq \theta \leq \pi$), $\varphi$ is azimuth angle ($0 \leq \theta < 2\pi$). Then polarization vectors will have the form:

$$\mathbf{e}_\perp = (\sin\varphi, -\cos\varphi, 0) \quad, \quad \mathbf{e}_\| = (-\cos\theta\cos\varphi, -\cos\theta\sin\varphi, \sin\theta) \quad. \tag{3.3}$$

Because of the symmetry of the problem the result of calculations will not depend on $\varphi$ and we may restrict to $\varphi = \pi$:

$$\mathbf{s} = (-\sin\theta, 0, \cos\theta) \quad, \quad \mathbf{e}_\perp = (0, 1, 0) \quad, \quad \mathbf{e}_\| = (\cos\theta, 0, \sin\theta) \quad. \tag{3.4}$$

Projecting in (3.1) on the coordinate axes, we then obtain the following system of coupled scalar equations:

$$f_j^x = E_j \cos\theta + \sum_{l \neq j} C_{lj}^x f_l^x \quad ,$$

$$f_j^y = E_j + \sum_{l \neq j} C_{lj}^y f_l^y \quad , \tag{3.5a}$$

$$f_j^z = E_j \sin\theta + \sum_{l \neq j} C_{lj}^z f_l^z \quad ,$$

with

$$C_{lj}^x = C_{lj}^y = \frac{\alpha}{a_0^3} \left[ \frac{k^2 a_0^2}{|l-j|} + \frac{ika_0}{(l-j)^2} - \frac{1}{|l-j|^3} \right] \exp\left(ika_0 |l-j|\right) \quad ,$$

$$C_{lj}^z = \frac{2\alpha}{a_0^3} \left[ -\frac{ika_0}{(l-j)^2} + \frac{1}{|l-j|^3} \right] \exp\left(ika_0 |l-j|\right) \quad , \tag{3.5b}$$

where $E_j = \exp\left[-ika_0(j-1)\cos\theta\right]$ corresponds to free field.

The equations (3.5) can be interpreted as follows: The atom $j$ undergoes the spontaneous emission. The influence on this by an other atom $l \neq j$ in the chain is given by the factors $C_{lj}$ of (3.5b). The problem contains three lengthscales: the interatomic distance $a_0$, the wavelength $\lambda = 2\pi/k$ of the mode in question and the polarizability $\alpha^{1/3}$. The physics is characterized by two dimensionless parameters $\alpha/a_0^3$ and $a_0 k$. The parameter $ka_0$ is assumed to be small because we are dealing with optical quantum transitions. The common factor $\alpha/a_0^3$ gives the strength of the interaction between the atoms. The phase of the different $C_{lj}$ is determined by the wavelength of the mode according to the distance $a_0|l-j|$ of the contributing atom $l$. The term proportional to $|l-j|^{-3}$ represent the Coulomb part of the interaction. It falls of quickly with the distance of the atom $l$ so that retardation effects do not play a role. The other terms are the retardation terms. With $ka_0 \ll 1$ their contribution will be small. The overall picture is therefore that the sums in (3.5a) will essentially reduce to only a few terms. We want to discuss this in detail.

### 3.1. Numerical results

To be able to judge the approximations we are going to made below, we make a numerical approach first. Equations (3.5) are solved with the iterative method of Zeidel [25]. For the wavelength $\lambda = 628 \ nm$ (corresponding to the resonance frequency $\omega_0$) we choose the interatomic distance $a_0$ to be $a_0 = \lambda/628$ so that $ka_0 \ll 1$ is fulfilled. To make the resulting figures easily interpretable we choose as number of atoms in the chain $N = 629$ so that its length becomes $L = \lambda$.

The resulting $y$- component of the mode functions is shown in Fig.1 for different values of the normalized polarizability $C = \alpha/a_0^3$. They demonstrate that there is no visible modification of the wave length of electromagnetic radiation. This means that the chain behaves like a dielectric medium with a refractive index equal to unity. The reason for the fact that phase properties of the field don't undergo perceptible changes is the following: According to the exponential factor in (3.5b) essential phase shifts at the position of the $j$-th atom in comparison with the phase of the free field may occur only due to the interactions of the $j$-th atom with far distant atoms, i.e., if the quantity $|l-j|$ is sufficiently large. However, for large $|l-j|$ the coefficients $C_{lj}$ tend to zero because of the square brackets in (3.5b). This fact explains the discovered property of the chain.

Because the amplitudes of the real and imaginary parts of the field in Fig. 1 a-c are equal to one another, the direction of propagation of the field in a one-dimensional chain coincides with the direction of propagation of the external field. This fact may apparently be considered as a consequence of the property discussed above.

We add a remark: The refractive index forms as a result of a summation of phase shifts produced by every sufficiently close atom. Therefore, owing to the character of the dipole-dipole interaction, there may well be a non-trivial refractive index for systems where the atoms are not arranged as a linear chain as discussed here. For higher dimensional arrays every atom has a larger number of neighbours. A more detailed discussion of this problem will be given in a subsequent paper.

The physical situation described above changes though a few interatomic distances away from the ends of the chain. There we observe according to Fig.1d rapid oscillations going back to the violation of the translational symmetry. The amplitudes of these oscillations and the distances from the ends of the chain at which they occur depend on the polarizability $\alpha$. With the increase of $\alpha$ the distances, at which boundary effects are essential, increase too.

To represent these oscillations which happen on top of the behaviour in the free space case ($\alpha = 0$, first term in (3.5a)) we make the ansatz

$$f_j^x = T_j^x E_j \cos\theta \quad , \quad f_j^y = T_j^y E_j \quad , \tag{3.6}$$

$$f_j^z = T_j^z E_j \sin\theta \quad , \quad T_j^x = T_j^y \quad , \quad j = 1, ..., N.$$

The amplitude factors $T_j^x$ and $T_j^z$ remain to be specified, $j$ denotes the position of emitting atom. Analytical expressions can be obtained on the basis of physically simplifying assumptions. We turn to this now.

### 3.2. Nearest neighbours approximation

If $\alpha$ is small enough we have shown that it is justified to take into account in the coupled equations (3.5) only the interactions with one or two nearest neighbours. This leads to analytical approximate solutions of (3.5) which will later be compared with the numerical results.

Restricting first to the interaction with only *one nearest neighbour*, equations (3.5) take the form:

$$f_j^x = E_j \cos\theta + C_1^x \left( f_{j-1}^x + f_{j+1}^x \right) \quad ,$$

$$f_j^y = E_j + C_1^y \left( f_{j-1}^y + f_{j+1}^y \right) \quad , \tag{3.7}$$

$$f_j^z = E_j \sin\theta + C_1^z \left( f_{j-1}^z + f_{j+1}^z \right) \quad ,$$

$$C_1^x = C_1^y = -\frac{\alpha}{a_0^3} \quad , \quad C_1^z = \frac{2\alpha}{a_0^3} = -2C_1^x \quad .$$

Here we have taken into account only the Coulomb interaction of the dipoles because of $ka_0 \ll 1$ in the optical range. We seek for the solution of the system (3.7) in the form

$$f_j^x = T_0^x E_j \cos\theta \quad , \quad f_j^y = T_0^y E_j \quad , \quad f_j^z = T_0^z E_j \sin\theta \quad , \tag{3.8a}$$

for the positions $2 \leq j \leq N-1$ and

$$f_j^x = T_1^x E_j \cos\theta \quad , \quad f_j^y = T_1^y E_j \quad , \quad f_j^z = T_1^z E_j \sin\theta \quad , \tag{3.8b}$$

for the ends of the chain $(j = 1, N)$. Substituting (3.8a), (3.8b) into (3.7) and using again $ka_0 \ll 1$, it is easy to obtain the unknown amplitudes. The result is

$$T_0^x = T_0^y = \frac{1}{1+2C} \quad , \quad T_0^z = -2T_0^x \quad , \tag{3.9}$$

$$T_1^x = T_1^y = \frac{1+C}{1+2C} \quad , \quad T_1^z = -2T_1^x \quad .$$

To improve the approximation we now take into account interactions with *two nearest neighbours*. In this case the system of equations (3.5) reduces to

$$f_j^x = E_j \cos\theta + C_1^x \left(f_{j-1}^x + f_{j+1}^x\right) + C_2^x \left(f_{j-2}^x + f_{j+2}^x\right) \quad ,$$

$$f_j^y = E_j + C_1^y \left(f_{j-1}^y + f_{j+1}^y\right) + C_2^y \left(f_{j-2}^y + f_{j+2}^y\right) \quad , \tag{3.10}$$

$$f_j^z = E_j \sin\theta + C_1^z \left(f_{j-1}^z + f_{j+1}^z\right) + C_2^z \left(f_{j-2}^z + f_{j+2}^z\right) \quad ,$$

$$C_2^x = C_2^y = -\frac{1}{8}\frac{\alpha}{a_0^3} \quad , \quad C_2^z = \frac{1}{4}\frac{\alpha}{a_0^3} = -2C_2^x \quad ,$$

with coefficients $C_1^x$, $C_1^y$, $C_1^z$ defined in (3.7). We look for a solution of the system (3.10) in the form (3.8a) for $3 \leq j \leq N-2$, in the form (3.8b) for $j = 1, N$, and in the form

$$f_j^x = T_2^x E_j \cos\theta \quad , \quad f_j^y = T_2^y E_j \quad , \quad f_j^z = T_2^z E_j \sin\theta \quad , \tag{3.11}$$

at the positions $j = 2, N-1$. Substituting (3.8), (3.11) into (3.10) we then obtain

$$T_0^x = T_0^y = \frac{1}{1+2.25C} \quad ,$$

$$T_2^x = T_2^y = \frac{1 + 0.125C - 2.125C^2}{(1+2.25C)(1-C^2)} \quad ,$$

$$T_1^x = T_1^y = 1 - C\frac{1.125 + 0.125C - 2C^2}{(1+2.25C)(1-C^2)} \quad , \tag{3.12}$$

$$T_l^z = -2T_l^x \quad , \quad l = 0, 1, 2 \quad .$$

The result for atoms at the ends of the chain and far from the ends ($T_0^{x,y,z}$) is given in Table 1 for different values of the normalized polarizability $C = \alpha/a_0^3$. Note that the disagreement between the numerical calculation and the calculation based on the nearest neighbour approximation is only of the order of a few percent. This approximation reflects therefore very well the underlying physical processes.

### 3.3. The infinite dipole chain

To complete the discussion we treat the infinite dipole chain. The limit $N \to \infty$ corresponds to the neglect of boundary effects. To obtain an analytical solutions of the system (3.5) we make the "ansatz" (3.8a) for every $j$. Substituting (3.8a) into (3.5) and omitting small terms of the order $ka_0$ and higher, we obtain the following expressions for the amplitudes

$$T_0^x = T_0^y = \frac{1}{1 + 2C\zeta(3)} \quad , \quad T_0^z = -\frac{2}{1 + 2C\zeta(3)} \quad , \tag{3.13}$$

where $\zeta(3) = 1.202$ is the third order Riemann $\zeta$-function. The results $T_0$ of (3.13) agree exactly with the numerical results of the last coulomn of Table 1 which refer to the inner atoms. Thus, due to predominant character of the Coulomb part of the dipole-dipole interaction, boundary effects in a linear chain don't influence the behaviour of the mode functions far from the ends of the chain.

## 4. LIFE TIME OF THE EXCITED STATE AND FREQUENCY SHIFT OF THE ATOMIC TRANSITION

We are now prepared to work out the life time of the excited state and the frequency shift of the atomic transition (Lamb shift) in a linear chain of atoms in making use of formula (2.6) and the results obtained in section 3. Here we shall not limit the treatment to nearest-neighbour approximation but use the general expression (3.6). Taking into account (3.3), the mode functions may be written as

$$\mathbf{f}_{\mathbf{k}\lambda}(\mathbf{r}_j) = \mathbf{E}_{\lambda j} \exp(ik\mathbf{s}\mathbf{r}_j) \quad ,$$

$$\mathbf{E}_{\parallel j} = \left(-T_j^x \cos\theta \cos\varphi, -T_j^x \cos\theta \sin\varphi, T_j^z \sin\theta\right) \quad , \tag{4.1}$$

$$\mathbf{E}_{\perp j} = T_j^x (\sin\varphi, -\cos\varphi, 0) \quad , \quad j = 1, \cdots, N \quad .$$

We substitute (4.1) into (2.6) and replace the mode summation by a summation over polarizations and an integration over wave vectors according to

$$\sum_{\mathbf{k}\lambda} \to \frac{V_R}{(2\pi c)^3} \int \omega_k^2 d\omega_k d\Omega \sum_\lambda \quad ,$$

where $d\Omega$ is a solid angle element in the direction of emission of spontaneous photon. Integrating over the frequencies we get

$$\frac{1}{\tau(\mathbf{r}_j)} = \frac{1}{\tau_1} \frac{3}{8\pi} \sum_\lambda \int_{4\pi} (\mathbf{u}_d \mathbf{E}_{\lambda j})^2 d\Omega \quad , \tag{4.2}$$

where $\tau_1$ is the life time of a single atom in free space. The unit vector $\mathbf{u}_d$ of (2.1) which defines the orientation of the transition dipole moment of the emitting atom at $\mathbf{r}_j$ may be written as

$$\mathbf{u}_d = (\sin\theta_d \cos\varphi_d, \sin\theta_d \sin\varphi_d, \cos\theta_d). \tag{4.3}$$

Substituting (4.1), (4.3) into (4.2) and integrating over the angles we finally obtain for the position dependent life time $\tau(\mathbf{r}_j)$

$$\frac{1}{\tau(\mathbf{r}_j)} = \frac{1}{\tau_1} \left[ \left(T_j^x\right)^2 \sin^2\theta_d + \left(T_j^z\right)^2 \cos^2\theta_d \right] \quad . \tag{4.4}$$

Corresponding calculations for the frequency shift of the atomic transition give an analogous result

$$\Delta(\mathbf{r}_j) = \Delta_1 \left[ \left(T_j^x\right)^2 \sin^2\theta_d + \left(T_j^z\right)^2 \cos^2\theta_d \right] \quad , \tag{4.5}$$

where $\Delta_1$ is the free space frequency shift. This is the intended result. For the different $T$ see (3.12) or Table 1.

The dependence of the life time $\tau(\mathbf{r}_j)$ on the number $j$ of the initially excited atom is shown in Fig. 2 for the case $\theta_d = \pi/2$. We see that the life time of the excited state may be considerably larger for atoms near the end of the chain as compared with an atom in free space. Only for atoms several interatomic distances away from the end, the value of the life time in an infinite chain is reached which is still larger than the one in free space. All these effects increase with increasing $C = \alpha/a_0^3$. Note that $\tau^{-1}(\mathbf{r}_j)$ and $\Delta(\mathbf{r}_j)$ depend in the same characteristic way on the angle $\theta_d$ which the transition dipole moment forms with the direction of the linear chain.

## 5. ANGULAR DISTRIBUTION OF THE SPONTANEOUSLY EMITTED PHOTONS

Finally we investigate the angular distribution of the photons spontaneously emitted by an atom in a linear chain of atoms. To do so we introduce the operator of the photon number in the $\mathbf{k}\lambda$-mode: $n_{\mathbf{k}\lambda} = c_{\mathbf{k}\lambda}^+ c_{\mathbf{k}\lambda}$. Going back to section 2 and the Hamiltonian (2.1) and calculating the corresponding commutators, we obtain the Heisenberg equation which governs the time evolution of the operator of the number of photons emitted by the $j$-th atom

$$\dot{n}_{\mathbf{k}\lambda}(\mathbf{r}_j, t) = i 2 \frac{\omega_0 d_0}{\hbar c} \left[ g_{\mathbf{k}\lambda}(\mathbf{r}_j) c_{\mathbf{k}\lambda}(t) - g_{\mathbf{k}\lambda}^*(\mathbf{r}_j) c_{\mathbf{k}\lambda}^+(t) \right] \quad . \tag{5.1}$$

The photon creation and annihilation operators are defined by the formula (2.6). After elementary transformations the vacuum mean value of the time derivative of the photon number operator may be wtitten in the form

$$\langle \dot{n}_{\mathbf{k}\lambda}(\mathbf{r}_j, t) \rangle = 2\pi \left( \frac{\omega_0 d_0}{\hbar c} \right)^2 |g_{\mathbf{k}\lambda}(\mathbf{r}_j)|^2 \times$$

$$\times \left[ \langle \sigma^+ \sigma \rangle_j \delta(\omega_k - \omega_0) + \langle \sigma \sigma^+ \rangle_j \delta(\omega_k + \omega_0) \right] \quad . \tag{5.2}$$

We study as usual the intensity in the direction of emission $\mathbf{s}$ (3.2) at the time $t$ for the atom located at the position $\mathbf{r}_j$:

$$I(\mathbf{r}_j, \mathbf{s}, t) = \frac{V_R}{(2\pi c)^3} \sum_\lambda \int \hbar \omega_k \langle \dot{n}_{\mathbf{k}\lambda}(\mathbf{r}_j, t) \rangle \omega_k^2 d\omega_k \quad . \tag{5.3}$$

Substituting (5.2) into (5.3) we get an expression which is similar to the one obtained in [23]

$$I(\mathbf{r}_j, \mathbf{s}, t) = \frac{\omega_0^4 d_0^2}{2\pi c^3} \langle \sigma^+ \sigma \rangle_j \sum_\lambda |\mathbf{u}_d \mathbf{f}_{\mathbf{k}_0 \lambda}(\mathbf{r}_j)|^2 \quad . \tag{5.4}$$

The $\mathbf{f}_{\mathbf{k}_0\lambda}(\mathbf{r}_j)$ are still unspecified. Therefore, the formula (5.4) is the general expression for the emission intensity and it may be used for the investigation of the angular distribution of spontaneously emitted photons in media with arbitrary atomic distribution. In the case of a single atom in free space we have $\mathbf{f}_{\mathbf{k}\lambda}(\mathbf{r}_j) = \mathbf{e}_\lambda \exp(ik\mathbf{s}_I\mathbf{r}_j)$ and the formula (5.4) gives the well-known result for the intensity of dipole radiation:

$$I_d(\mathbf{s}, t) = \frac{\omega_0^4 d_0^2}{2\pi c^3} \langle \sigma^+ \sigma \rangle \left[1 - (\mathbf{u}_d \mathbf{s})^2\right] \quad . \tag{5.5}$$

To discuss the linear chain, we define a new unit vector

$$\mathbf{u}'_d(\mathbf{r}_j) = \left[\sin^2 \theta_d + \left(\frac{T_j^z}{T_j^x}\right)^2 \cos^2 \theta_d\right]^{-1/2} \left(\sin \theta_d \cos \varphi_d, \sin \theta_d \sin \varphi_d, \frac{T_j^z}{T_j^x} \cos \theta_d\right) \quad . \tag{5.6}$$

Note that the direction of $\mathbf{u}'_d$ depends on $T_j^z/T_j^x$ and therefore on the position of the atom in the chain. Making use of this notation and of the formulas (3.3) and (4.1), we obtain in this case finally

$$I(\mathbf{r}_j, \mathbf{s}, t) = \frac{\omega_0^4 d_0^2}{2\pi c^3} \langle \sigma^+ \sigma \rangle_j \left(T_j^x\right)^2 \left[1 - (\mathbf{u}'_d \mathbf{s})^2\right] \quad . \tag{5.7}$$

As a result it follows from (5.7) that the characteristic structure of the angular distribution of the photons emitted by the atom embedded in a linear chain is analogous to the one of a single atom in free space [equation (5.5)]. The dipole-dipole interaction leads to two modifications: There is a factor $\left(T_j^x\right)^2$ in (5.7) which diminishes the intensity uniformly for all directions. Secondly the well-known free-space angular distribution is rotated by an angle $\gamma$ given by

$$\cos \gamma = \mathbf{u}_d \mathbf{u}'_d \quad . \tag{5.8}$$

In the optical range we have $T_j^z = -2T_j^x$ so that the angle $\gamma$ becomes the same for all atoms. Substituting (4.3), (5.6) into (5.8) we get

$$\cos \gamma = \frac{1 - 3\cos^2 \theta_d}{\sqrt{1 + 3\cos^2 \theta_d}} \quad . \tag{5.9}$$

The dependence of the rotation angle $\gamma$ on the orientation of the transition dipole moment is shown in Fig.3.

## 6. CONCLUDING REMARKS

We have studied the single-atom spontaneous emission in a linear chain of two-level atoms. Particular attention has been payed to emitting atoms situated very close to the ends of the chain. It is shown that the respective life time, Lamb shift and the angular distribution of the intensity of the spontaneously emitted photons change considerably as compared with atoms in the middle of the chain or atoms in free space.

To treat the quantized electromagnetic field we did not rely on the macroscopic Maxwell equations with boundary conditions and space dependent refractive index, but used the integro-differential equations for the local field instead. From the point of view of QED the respective integro-differential equations for the local field take into account that the emission of a real photon may be accompagnied with an exchange of virtual photons with the other atoms of the chain. The discrete structure of the medium and the physical properties of the single atoms at given positions are thereby naturally represented. The fundamental interatomic interaction is the dipole-dipole interaction. We remark that a chain of dipoles also appears in the problem of quantum optics in the three-dimensional microscopic cavities, compare [26,27]. However, the behaviour of these image dipoles does not agree with the one of the real dipoles which build up the medium in our case.

## ACKNOWLEDGMENTS

This work has been supported by the Deutsche Forschungsgemeinschaft and the Optik Zentrum Konstanz. One of us (K.V.K.) would like to thank the members of the AG Audretsch at the University of Konstanz for many discussions and kind hospitality.16

The dependence of the rotation angle $\gamma$ on the orientation of the transition dipole moment is shown in Fig.3.

## 6. CONCLUDING REMARKS

We have studied the single-atom spontaneous emission in a linear chain of two-level atoms. Particular attention has been payed to emitting atoms situated very close to the ends of the chain. It is shown that the respective life time, Lamb shift and the angular distribution of the intensity of the spontaneously emitted photons change considerably as compared with atoms in the middle of the chain or atoms in free space.

To treat the quantized electromagnetic field we did not rely on the macroscopic Maxwell equations with boundary conditions and space dependent refractive index, but used the integro-differential equations for the local field instead. From the point of view of QED the respective integro-differential equations for the local field take into account that the emission of a real photon may be accompagnied with an exchange of virtual photons with the other atoms of the chain. The discrete structure of the medium and the physical properties of the single atoms at given positions are thereby naturally represented. The fundamental interatomic interaction is the dipole-dipole interaction. We remark that a chain of dipoles also appears in the problem of quantum optics in the three-dimensional microscopic cavities, compare [26,27]. However, the behaviour of these image dipoles does not agree with the one of the real dipoles which build up the medium in our case.

## ACKNOWLEDGMENTS

This work has been supported by the Deutsche Forschungsgemeinschaft and the Optik Zentrum Konstanz. One of us (K.V.K.) would like to thank the members of the AG Audretsch at the University of Konstanz for many discussions and kind hospitality.
16

# REFERENCES

[1] H. - I. Yoo and J. H. Eberly, Phys.Repts. **118**, 241 (1985)

[2] G. Barton and N. S. J. Fawcett, Phys.Repts. **170**, 1 (1988)

[3] A. J. Campillo, J. H. Eversole and H. - B. Lin, Mod.Phys.Lett.B **6**, 447 (1992)

[4] P. Meystre, Phys.Repts. **219**, 243 (1992)

[5] H. Walther, Phys.Repts. **219**, 263 (1992)

[6] V. P. Klochkov, Optika i spectroscopiya **74**, 676 (1993), in russian

[7] A. N. Oraevsky, Uspekhi Fizicheskikh Nauk **164**, 415 (1994), in russian

[8] H. Khosravi and R. Loudon, Proc.Roy.Soc.Lond.A, **433**, 337 (1991)

[9] M. Janowicz and W. Zakowicz, Phys.Rev.A, **50**, 4350 (1994)

[10] O.N.Gadomsky and K.V.Krutitsky, J.Appl.Spectr. **63**, 278 (1996), in russian

[11] O.N.Gadomsky and K.V.Krutitsky, JOSA B **13**, 1676 (1996)

[12] K.V.Krutitsky and S.V.Suhov, J.Phys.B **30**, 5341 (1997)

[13] O.N.Gadomsky and K.V.Krutitsky, Zh.Eksp.Teor.Fiz. **106**, 936 (1994) [JETP **79**, 513 (1994)]

[14] O.N.Gadomsky and K.V.Krutitsky, Proc.SPIE **2799**, 77 (1996)

[15] M. Born and E. Wolf, Principles of optics (Pergamon, New York, 1968), Chap.2

[16] L.Rosenfeld, Theory of electrons (North Holland Publishing Company, Amsterdam, 1951), Chap.6

[17] F.C.Spano and J.Knoester, Adv.Magn.Opt.Res. **18**, 117 (1994)

[18] V.Malyshev and P.Moreno, Phys.Rev.A **53**, 416 (1996)

[19] A.I.Zaitsev, V.A.Malyshev and E.D.Trifonov, Zh.Eksp.Teor.Fiz. **84**, 475 (1983) [Sov.Phys.JETP **57**, 275 (1983)]

[20] O.N.Gadomsky and K.V.Krutitsky, Quant.Semicl.Opt. **9**, 343 (1996)

[21] R.J.Glauber and M.Lewenstein, Phys.Rev.A **43**, 467 (1991)

[22] G.Björk, S.Machida, Y.Yamamoto and K.Igeta, Phys.Rev.A **44**, 669 (1991)

[23] L. Allen and J. H. Eberly, Optical resonance and two-level atoms (Wiley-Interscience, New York, 1975)

[24] W. Heitler, Quantum theory of radiation (Clarendon, Oxford, 1954)

[25] A. A. Amosov, Yu. A. Dubinsky and N. V. Kopchenova, Numerical methods for engineers (Vyshaya shkola, Moscow, 1994), Chap.6, in russian

[26] P. W. Milonni and P. L. Knight, Opt.Commun. **9**, 119 (1973)

[27] J. P. Dowling, M. O. Scully and F. DeMartini, Opt.Commun. **82**, 415 (1991)

**FIGURE CAPTIONS**

Fig.1. Numerical results for the position dependence of the mode function in a linear atomic chain extending from $z = 0$ to $z = -628\ nm$. The interatomic distance $a_0$ is $1\ nm$. $C = \alpha/a_0^3$ with the atomic polarizability $\alpha$. The wavelength of the free space photon field is assumed to be equal to the length of the chain. It can be read off from the Fig.a-c that in a chain the resulting wavelength is the same. There are strong oscillations a few interatomic distances away from the ends of the chain which are shown in detail in Fig.1d. The solid and the dashed line correspond to the real and imaginary part, respectively.

Fig. 2. Life time $\tau$ of a single excited atom in a linear chain. $j$ denotes the number of the position of the atom counting from the end of the chain. $\tau_1$ is the life time in free space. For the atomic transition the wave length $\lambda_0 = 628\ nm$ and the orientation of transition dipole moment was assumed to be perpendicular to the direction of the chain ($\theta_d = \pi/2$). The other parameters are chosen as for Fig.1.

Fig.3. For an atom in a chain the free space distribution of emitted photons is rotated by an angle $\gamma$ which depends on the orientation of the transition dipole moment (angle $\theta_d$) relative to the direction of the chain. The corresponding dependence is shown in this figure.

Table 1. Amplitude factors of the mode functions obtained by approximations and numerical calculations.

| C | interaction with one nearest neighbor | | | interaction with two nearest neighbors | | | numerical solution of the system (3.5) | | |
|---|---|---|---|---|---|---|---|---|---|
|   | $T_1^x$ | $T_2^x$ | $T_0^x$ | $T_1^x$ | $T_2^x$ | $T_0^x$ | $T_1^x$ | $T_2^x$ | $T_0^x$ |
| 0.1 | 0.917 | —— | 0.833 | 0.908 | 0.817 | 0.816 | 0.902 | 0.812 | 0.806 |
| 0.2 | 0.857 | —— | 0.714 | 0.846 | 0.675 | 0.690 | 0.839 | 0.668 | 0.675 |
| 0.3 | 0.813 | —— | 0.625 | 0.807 | 0.555 | 0.597 | 0.800 | 0.546 | 0.581 |
| 0.4 | 0.778 | —— | 0.556 | 0.786 | 0.445 | 0.526 | 0.787 | 0.425 | 0.510 |
| 0.5 | 0.750 | —— | 0.500 | 0.784 | 0.333 | 0.471 | 0.817 | 0.262 | 0.454 |